\documentclass[aps,pra,twocolumn,superscriptaddress,noeprint,longbibliography, nofootinbib,floatfix]{revtex4-2}
\usepackage{graphicx}% Include figure files
\usepackage{bm}% bold math
\usepackage{amsmath, amssymb}
\usepackage{booktabs}
\usepackage{mathrsfs}
\usepackage{adjustbox}
\usepackage{makecell}
\usepackage{float}
\usepackage[normalem]{ulem}
\usepackage{pifont}
\usepackage{placeins}
\usepackage[mathscr]{euscript}
\usepackage{physics}
\usepackage{xcolor}
\usepackage{siunitx}
\usepackage{comment}
\usepackage{appendix}
\usepackage{mathtools}
\usepackage[%
  pdfstartview=FitH,%
  breaklinks=true,%
  bookmarks=true,%
  colorlinks=true,%
  anchorcolor=black,%
  citecolor=blue,
  filecolor=black,%
  menucolor=black,%
  urlcolor=blue,%
  linkcolor=blue,%
 ]{hyperref}
\usepackage[normalem]{ulem}
\newcommand{\stkout}[1]{\ifmmode\text{\sout{\ensuremath{#1}}}\else\sout{#1}\fi}
\graphicspath{ {./Figures/} }

\def\equationautorefname~#1\null{Eq. (#1)\null}

\newcommand{\appref}[1]{\hyperref[#1]{App.~\ref*{#1}}}
\usepackage[all]{hypcap} %
\usepackage{physics}

\usepackage[backgroundcolor=green!40,textsize=tiny]{todonotes}
\begin{document}
\title{Antiscarring from eigenstate stacking in a chaotic spinor condensate}

\author{Zhongling~Lu}
\affiliation{Department of Chemistry and Chemical Biology, Harvard University, Cambridge,
Massachusetts 02138, USA}
\affiliation{Zhiyuan College, Shanghai Jiao Tong University, Shanghai 200240, China}

\author{Anton M.~Graf}
\affiliation{Harvard John A. Paulson School of Engineering and Applied Sciences,
Harvard, Cambridge, Massachusetts 02138, USA}
\affiliation{Department of Chemistry and Chemical Biology, Harvard University, Cambridge,
Massachusetts 02138, USA}

\author{Eric J.~Heller} 
\affiliation{Department of Chemistry and Chemical Biology, Harvard University, Cambridge,
Massachusetts 02138, USA}
\affiliation{Department of Physics, Harvard University, Cambridge, Massachusetts 02138, USA}

\author{Joonas~Keski-Rahkonen}
\affiliation{Department of Chemistry and Chemical Biology, Harvard University, Cambridge,
Massachusetts 02138, USA}
\affiliation{Department of Physics, Harvard University, Cambridge, Massachusetts 02138, USA}

\author{Ceren~B.~Dag}
\email{ceren_dag@g.harvard.edu}
\affiliation{Department of Physics, Harvard University, Cambridge, Massachusetts 02138, USA}
\affiliation{ITAMP, Center for Astrophysics, Harvard $\&$ Smithsonian, Cambridge, Massachusetts 02138, USA}
\affiliation{Department of Physics, Indiana University, Bloomington, Indiana 47405, USA}

%%%%%%%%%%%%%%%%%%%%%%%%%%%%%%%%%%%%%%%%%%%%%%%%%%%%
\begin{abstract}
\noindent
We reveal a feature of quantum scarring in systems with many particles: Quantum scars, living densely near an unstable periodic orbit, must be compensated by corresponding antiscarred states suppressed there to establish the uniformity of the whole. The uniformity of the underlying phase space is linked to early-time dynamics --- a regime beyond the predictions of random matrix theory and encapsulated in the eigenstate stacking theorem. By extending the domain of the stacking theorem, we apply our theory to a chaotic spinor Bose-Einstein condensate, whose quantum scar dynamics have recently been observed in the laboratory. Our work uncovers how scarring of some eigenstates affects the rest of the chaotic and thermal spectrum in quantum systems with many particles.
\\

\noindent \text{DOI: \href{https://link.aps.org/doi/10.1103/k594-zxlj}{10.1103/k594-zxlj}}

\end{abstract}

%%%%%%%%%%%%%%%%%%%%%%%%%%%%%%%%%%%%%%%%%%%%%%%%%%%%

\maketitle

\section{Introduction}

%\noindent
Scars are quantum eigenstates that exhibit enhanced probability density around unstable periodic orbits (UPOs) that reside in an underlying classical chaotic phase space~\cite{Heller_book_2018} --- a phenomenon that is an archetypal example of quantum-classical correspondence and is responsible for an important correction to a simple ergodicity assumption based on classical chaos or the alike notions in quantum chaology, e.g.,~Berry's random wave conjecture \cite{berry1977regular}, ergodicity theorems \cite{shnirel1974ergodic,zelditch1987uniform} and Gutzwiller's trace formula \cite{gutzwiller1971periodic}. First identified in quantum billiards having fully chaotic classical dynamics, quantum scarring marks its 40$^{\rm th}$ anniversary \cite{heller1984bound}, and continues to advance our understanding of quantum chaos today in single-particle \cite{bogomolny1988smoothed, de1994scars, vergini2000semiclassical,joonas2017scar,joonas2019lissajous, keski-rahkonen_j.phys.conden.matter_31_105301_2019, BB_article, chalangari2025variational, keskirahkonen_phys.rev.e_112_L012201_2025, 
mondal2020chaos,pilatowsky2021ubiquitous} and many-body systems \cite{lukin2017-51atom,Serbyn_2021,ho2019pxp,PhysRevX.10.011055,Hallam2023,PhysRevLett.130.250402,evrard-QuantumScars-2024,evrard-QuantumManybody-2024,ermakov2024periodic,pizzi2024scar,wang2023embedding,shen2024enhanced,garcia2025lindblad}. These efforts have led to abundant experimental evidence~\cite{crook2003scar-image,cabosart2017recurrent}, including recent experiments in graphene quantum dots \cite{ge2024direct} and a chaotic spin$-1$ Bose-Einstein condensate \cite{austin2024observation}. 

A defining characteristic of quantum scars is that they coexist with chaotic and thermal eigenstates, described by the random matrix theory (RMT) \cite{wigner1967random,brody1981random,mehta2004random,guhr1998random,Stockmann_1999} and the eigenstate thermalization hypothesis (ETH) \cite{srednicki1994chaos,deutsch1991quantum,Deutsch_2018,d2016quantum}. RMT captures the universal properties of quantum systems in energy windows smaller than the Thouless energy \cite{thouless1972numerical}, which corresponds to the onset of RMT at the Thouless time \cite{kos-ManyBodyQuantum-2018,PhysRevLett.121.060601,gharibyan-OnsetRandom-2018}. Subsequently, only the spectral correlations that develop at late times 
are often considered universal \cite{PhysRevB.99.174313,PhysRevX.10.041017,PhysRevE.106.024208}, although recent arguments on universality beyond the RMT exist \cite{PhysRevLett.121.060601,gharibyan-OnsetRandom-2018,dag-ManybodyQuantum-2023}. On the other hand, by investigating the probability density of all eigenstates in a sufficiently large energy window, i.e.,~a ``stack'', a universal feature, independent of any microscopic details, arises: the stack must be \textit{uniform} in phase space -- encapsulated by the \textit{eigenstate stacking theorem}~\cite{keski-rahkonen-AntiscarringChaotic-2024}. Here, the ``phase space'' of a quantum system can be thought of as an asymptotically exact analog of a classical phase space, e.g.,~the Husimi representation, essentially a coherent state projection. Specifically, we mean by uniformity in phase space over an energy interval, a density operator defined as $\rho_{E_0, \Delta E} = \int_{E_0-\Delta E/2}^{E_0+\Delta E/2}\langle \zeta \vert\delta (E- \hat H )\vert \zeta \rangle\ dE$ approaching a uniform distribution in a phase space for a Hamiltonian $\hat H$ within a measure basis $\vert \zeta \rangle$ \cite{Heller_book_2018}.

The uniformity can be understood by constraining the dynamics to early timescales, where ``early'' is defined as a timescale shorter than the period of the shortest periodic orbit (PO) in the underlying classical system. Before then, an initial state $\vert\phi(0)\rangle$ launched along the periodic orbit has a survival probability $\langle \phi(0)\vert \phi(t)\rangle $ that decays in a timescale shorter 
than the period of the orbit, and this timescale depends on the initial state. The decay holds until the return period of the orbit, and the Fourier transform of the decay, cut off before one period, leads to a smooth and non-undulating energy window, which is independent of the cutoff time. In consequence, the sum of the probability densities of all eigenstates in any representation, e.g.,~position, momentum or Husimi, weighted by that energy window function must agree with a classical ergodic distribution of the corresponding projection. 

For instance, in the case of a two-dimensional billiard, the ergodic coordinate space distribution is uniform with density $\rho = 1/A$ where $A$ is the area of the billiard. The distribution is best probed directly in phase space around a periodic orbit using Husimi projections, which are also employed here. If a class of eigenstates in the energy window has enhanced probability around stable or unstable POs, due to integrability or scarring, respectively, other eigenstates must necessarily be suppressed there to establish the uniformity of the whole, hence respecting the stacking theorem. In the case of quantum scarring and UPOs, this behavior is coined as \textit{antiscarring}~\cite{kaplan-ScarAntiscar-1999,keski-rahkonen-AntiscarringChaotic-2024}.

In this work, we show that chaotic spinor condensates that have genuine quantum scarring \cite{evrard-QuantumScars-2024,austin2024observation}, also exhibit antiscarring predicted by the stacking theorem. The model we apply for chaotic spinor condensates has a semiclassical limit based on the coherent state basis, which opens an avenue to investigate the quantum-classical correspondence in chaotic many-particle quantum systems, such as the effect of POs on the spectrum and eigenstates. We compute spectral rigidity \cite{dyson1963statistical} and the connected spectral form factor (SFF) \cite{guhr1998random,mehta2004random} to estimate the shortest PO frequency, confirming that the energy window to observe uniformity is indeed lower-bounded by the shortest PO frequency. 

While the perfect uniformity of the phase space is attainable in the thermodynamic limit $N \rightarrow \infty$, we observe approximate uniformity for finite-size condensates due to the properties of coherent states. Hence, our work extends the applicability of stacking theorem \cite{keski-rahkonen-AntiscarringChaotic-2024} to the case of generic measure states, e.g.,~states with defects such as coherent states. Importantly, despite the approximate uniformity in phase space, the antiscarring is still present even for small condensates. In this sense, our work highlights the implications of scarring of some eigenstates for the rest of the chaotic and thermal spectrum in quantum systems with many particles. 

%%%%%%%%%%%%%%%%%%%%%%%%%%%%%%%%%%%%%%%%%%%%%%%%%%%%
\section{Spin$-1$ chaotic spinor condensate}

\begin{figure}
    \centering
	\includegraphics[width=\columnwidth]{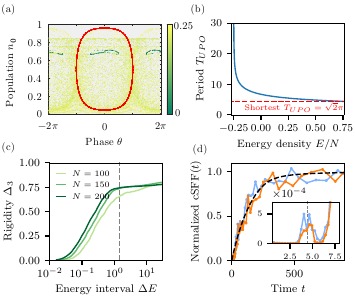}
    \caption{(a) The Poincare section at energy density $E=0.24$ with colorbar denoting the Lyapunov exponents of the trajectories. Almost entire phase space is chaotic, with a small regular island at $n_0\sim 0.7$. The periodic orbit highlighted in red is unstable with Lyapunov exponent $\lambda=0.31$. (b) Periods of the UPO family $T_{\text{UPO}}$ are found to be continuous. At $p=0.5$, $T_{\text{UPO}}$ spans from $\sqrt{2}\pi$ to infinity. (c) Spectral rigidity $\Delta_3$ as a function of energy window $\Delta E$ with increasing system size $N$ (light to dark green).    The saturation is observed after the energy width $\sim 2\pi/T_{\text{UPO}}^*$, marked by the dashed gray line. (d) Normalized connected spectral form factor (cSFF) computed for $N=100$ (blue) and $N=150$ (orange). The cSFF follows the GOE prediction marked with the dashed black curve, whereas at early times it peaks at the period of the shortest UPO $T_{\text{UPO}}^*$ (inset). }
	\label{fig:upo_period&poincare}
\end{figure}

\begin{figure}
    \centering
	\includegraphics[width=\columnwidth]{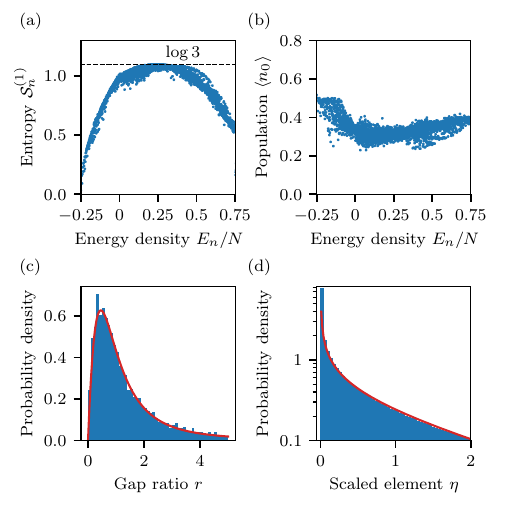}
    \caption{(a,b) One-body entanglement entropy and eigenstate expectation value $\langle n_0 \rangle$ with respect to energy density $E/N$ for $N=200$ atoms and $p=0.5$. (c) Distribution of ratio of nearest-neighbor energy levels (gap ratio) matches well with the Wigner-Dyson statistics of GOE (red). (d) Distribution of scaled eigenstate element $\eta$ within energy range $0.18 < E_n /N <0.3$ agrees with Porter-Thomas distribution of GOE (i.e.,~$\chi^2$ distribution with one degree of freedom), marked by the red line. }
	\label{fig:chaos_diagnosis}
\end{figure}

\begin{figure*}[htb!]
\centering
\includegraphics[width=\textwidth]{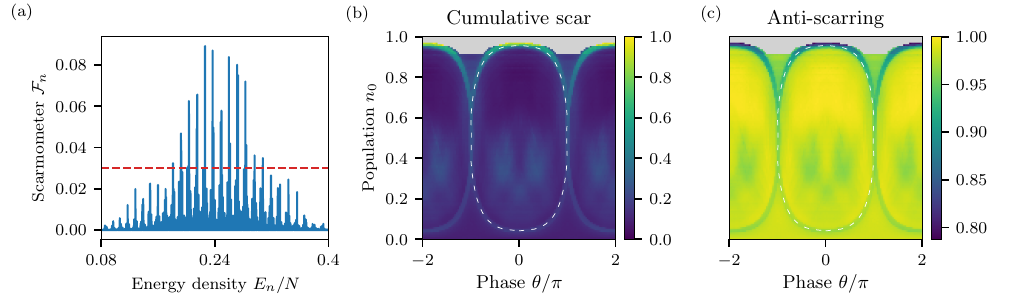}
\caption{ (a) The scarmometer $\mathcal{F}_n$ of \autoref{eq:scarmometer}, plotted with respect to energy density.  (b,c) Projected Husimi-Q distribution of cumulative scar and antiscarring around UPO at $E_0/N=0.24$ with $N=100$, \autoref{eq:proj_stack}. Here the colormap is normalized by the maximum value. The gray area indicates no density of states. The cumulative scar is obtained by stacking scar eigenstates with $\mathcal{F}_n > 0.03$ indicated by the dashed red line in (a). }
\label{fig:scar_antiscar}
\end{figure*}

\subsection{Model}

We consider a spin$-1$ Bose-Einstein condensate composed of $N$ bosonic atoms~\cite{pethick2008bose, stamper2013spinor}. While such a physical system is fundamentally a many-body system, here we assume a tight optical trap resulting in the decoupling of spatial and spin degrees of freedom such that the condensation occurs in a single spatial orbital leading to an all-to-all coupled spin model \cite{ohmi1998bose,law1998quantum,ho1998spinor,yi2002single,yang2019observation,evrard2021many,austin2024observation} of 
\begin{align}
		\hat H = \frac{c_1}{N}\left[\hat N_0(N-\hat N_0)+\frac{1}{2}(\hat N_+-\hat N_-)^2\right]+p(\hat W_++\hat W_-). \label{eq.Hamiltonian2ndQuant}
\end{align}
where $\hat N_m=\hat a_m^\dagger\hat a_m$, with $\hat a_m$ ($\hat a_m^{\dagger}$) the annihilation (creation) operator for the spin mode $m=0,\pm 1$. We note that the single-spatial mode approximation is achievable in experiments \cite{PhysRevLett.102.125301,yang2019observation,evrard2021many,austin2024observation}. The interaction Hamiltonian with strength $c_1/N$ originates from all-to-all Heisenberg interaction between atoms \cite{kawaguchi2012spinor}. 

In the following, we set $c_1=1$. 
The term $\hat W_{\pm}=\frac{1}{\sqrt{2}}\left(\hat a_\pm^\dagger\hat a_0 + \textrm{H.c.}\right)$ is the mode mixing term, which breaks the integrability of the interaction Hamiltonian \cite{evrard2021many,rautenberg2020classical} with strength $p$ referring to the Larmor frequency associated with the rotating in-plane field applied to the condensate~\cite{evrard-QuantumScars-2024}. The single-spatial mode approximation simplifies the model by reducing the dimension of the Hilbert space from exponential to quadratic in atom number. The system defined by the Hamiltonian in Eq.~\eqref{eq.Hamiltonian2ndQuant} has been theoretically shown~\cite{evrard-QuantumScars-2024} and experimentally~\cite{austin2024observation} verified with sodium atoms to exhibit a mixed spectrum, associated with a mixed phase space in its semiclassical limit, with both thermal and nonthermal eigenstates at small $p = 0.05$. Remarkably, this simple model hosts both types of scarring phenomena: quantum scars originating from UPOs~\cite{heller1984bound} as well as quantum many-body scar-like regular states originating from stable periodic orbits \cite{evrard-QuantumManybody-2024}. 

\subsection{Semiclassical limit}

In this work, we instead focus on a large $p= 0.5$ that consequently removes most of the regular regions in the classical phase space. This phase space is defined with respect to SU(3) symmetric coherent states $\ket{\zeta} = \frac{1}{\sqrt{N!}}[\sum_m \zeta_m\hat a^\dagger_m]^N\ket{0}$
with $\zeta_m=\sqrt{n_m} e^{i\phi_m}$, where $n_m\equiv N_m/N$. By demanding $\sum_m n_m = 1$ and a trivial global phase $\phi_0=0$, we parametrize the coherent states by four real numbers, $n_0$, $\theta=\phi_++\phi_-$, $m=n_+-n_-$ and $\eta=\phi_+-\phi_-$ \cite{evrard-QuantumScars-2024} which leads to classical equations of motion and Poincare sections, valid in the thermodynamic limit $N\rightarrow \infty$, and formally
\begin{eqnarray}
\dot n_0 &= & \hspace{1mm} p\sqrt{2n_0}\bigg [\sqrt{n_+}\sin \phi_+ + \sqrt{n_-}\sin\phi_- \bigg ]\,,\notag\\
\dot \theta &=& \hspace{1mm}2(1-2n_0) +p\bigg [\frac{2n_+-n_0}{\sqrt{2n_0n_+}}\cos\phi_+ \notag \\
&+&\frac{2n_--n_0}{\sqrt{2n_0n_-}}\cos\phi_-\bigg ]\,,\notag\\
\dot m &=& \hspace{1mm}p\sqrt{2n_0}\bigg [-\sqrt{n_+}\sin\phi_++\sqrt{n_-}\sin\phi_- \bigg]\,,\notag\\
\dot\eta &=& \hspace{1mm}-2m-p\sqrt{\frac{n_0}{2}}\bigg [\frac{\cos\phi_+}{\sqrt{n_+}}-\frac{\cos\phi_-}{\sqrt{n_-}}\bigg ]\,,\label{eq:eqnMotion}
\end{eqnarray}
where $\theta=\phi_++\phi_-$, $m=n_+-n_-$ and $\eta=\phi_+-\phi_-$. 
The mean-field energy per atom is
\begin{eqnarray}
E &=& [n_0(1-n_0)+\frac{m^2}{2}] \notag \\
&+&p\sqrt{2n_0}\left(\sqrt{n_+}\cos\phi_+ +\sqrt{n_-}\cos\phi_- \right)\,.\label{eq.meanfieldEnergy}
\end{eqnarray}
The phase space defined by these equations of motion is chaotic: \autoref{fig:upo_period&poincare}(a) shows a Poincare section at the plane $m=\eta=0$ where the color denotes the Lyapunov exponents of the classical trajectories. We observe an unstable periodic orbit (UPO) embedded in the chaotic phase space in \autoref{fig:upo_period&poincare}(a). There is in fact a family of UPO lying on plane $m=0,\eta=0$ with energy
\begin{equation}
\begin{aligned}
& E = n(1-n) + 2p\sqrt{n(1-n)}\cos{\theta/2},\label{eq:UPOenergy} \\
\end{aligned} 
\end{equation}
and determined by the following equations of motion,
\begin{equation}
\begin{aligned}
    \dot{n} &= 2p \sqrt{n(1-n)}\sin{(\theta/2)}, \\
    \dot{\theta} &= 2(1-2n) + 2p \frac{1-2n}{\sqrt{n(1-n)}} \cos{(\theta/2)}. \label{eq:EOMUPO}
\end{aligned}
\end{equation}
From Eqs.~\eqref{eq:UPOenergy} and~\eqref{eq:EOMUPO},  we derive the period of the UPOs $T_{\rm UPO}$ as,
\begin{eqnarray}
T_{\text{UPO}} =  \frac{4}{\sqrt{y_1-y_2}} K\left(\sqrt{\frac{y_1}{y_1-y_2}}\right), \label{eq:TUPO}
\end{eqnarray}
where $y_{1,2} = -(2p^2+E-\frac{1}{4})\pm 2p \sqrt{p^2+E}$ and $K(\cdot)$ is the complete elliptic integral of the first kind (see \appref{AppendixSec:UPOPeriod} for the derivation). For $p=0.5$, the shortest UPO period appears at $y_1=0,E=0.75$ leading to $T_\text{UPO}^{*}=\sqrt{2}\pi \simeq 4.44$. We also notice the divergence of UPO periods near $E=-0.25$, evident in \autoref{fig:upo_period&poincare}(b).

\subsection{Spectral and eigenstate properties}

Consistently with a chaotic phase space, the model determined by Eq.~\eqref{eq.Hamiltonian2ndQuant} bears several signatures of quantum chaos. For example, the gap ratio $r_n = \frac{E_{n+1}-E_n}{E_n - E_{n-1}}$ quantifies the level statistics and offers a direct comparison with RMT that does not depend on local density of states. We observe excellent agreement of $r_n$ with the Gaussian orthogonal ensemble (GOE) prediction in \autoref{fig:chaos_diagnosis}(c). Furthermore, the system obeys ETH in the strong sense, i.e.,~the spin$-0$ atom population $\langle n_0 \rangle=\langle \psi_n | \hat n_0 | \psi_n\rangle$ shows a uniform distribution with energy around our choice of energy density $E/N=0.24$, as observable in \autoref{fig:chaos_diagnosis}(b). 
Consistently, the particle entanglement entropy $ \mathcal{S}_n^{(1)}= -\operatorname{Tr}\left[\rho_n^{(1)} \ln \rho_n^{(1)}\right]$ of the one-body density matrix $\left[\rho_n^{(1)}\right]_{j k}=\left\langle\psi_n\right| \hat{a}_j^{\dagger} \hat{a}_k\left|\psi_n\right\rangle$, which quantifies the entanglement of one atom with the rest of the ensemble \cite{evrard2021observation}, exhibits the largest allowed value, $\mathcal{S}_n^{(1)} = \rm log(3)$, for these thermal states, \autoref{fig:chaos_diagnosis}(a). We also compute the eigenstate statistics within the energy window $0.18<E_n/N <0.3$, $\eta= D|c^i_n|^2$ in $\ket{\psi_n} = \sum_{i=1}^D c^i_n \ket{\phi_i}$, where $\ket{\phi_i}$ are the basis vectors and $D$ is the Hilbert space dimension. We find that eigenstate statistics follows perfectly the Porter-Thomas distribution of the eigenstates of the GOE matrices, $P(\eta)=\frac{1}{\sqrt{2\pi \eta}}\exp (-\eta/2)$ as seen in \autoref{fig:chaos_diagnosis}(d) which further confirms that the system has maximal quantum chaos. We underline that such maximal chaos is absent in the model with $p=0.05$. Despite the dominance of chaotic and thermal eigenstates in the spectrum, quantum scars still exist at $p=0.5$. 

\subsection{Quantum scars}

To visualize quantum scars, a standard way is to compute the Husimi-Q distribution $Q_n (\zeta) = \left| \braket{\zeta}{\psi_n} \right|^2$ of the quantum scar eigenstate $\ket{\psi_n}$ with measure states that span the phase space in the classical limit, e.g.,~coherent states $\ket{\zeta}$ in our case \cite{evrard-QuantumScars-2024}. Subsequently, a degree of scarness called scarmometer~\cite{keski-rahkonen-AntiscarringChaotic-2024, evrard-QuantumScars-2024} can be defined as 
\begin{equation}\label{eq:scarmometer}
\mathcal{F}_n = \oint_{\text {UPO}} d \zeta \left|\braket{\psi_n}{\zeta}\right|^2  =  \oint_{\text {UPO}} d \zeta Q_n(\zeta),
\end{equation}
which measures the overlap between an eigenstate $\ket{\psi_n}$ and the measure states $\ket{\zeta}$ that are on the UPO. \autoref{fig:scar_antiscar}(b) illustrates the eigenstates that are scarred by the UPO at  $E_n/N=0.24$. 

\begin{figure*}[htb!]
    \centering
	\includegraphics[width=\textwidth]{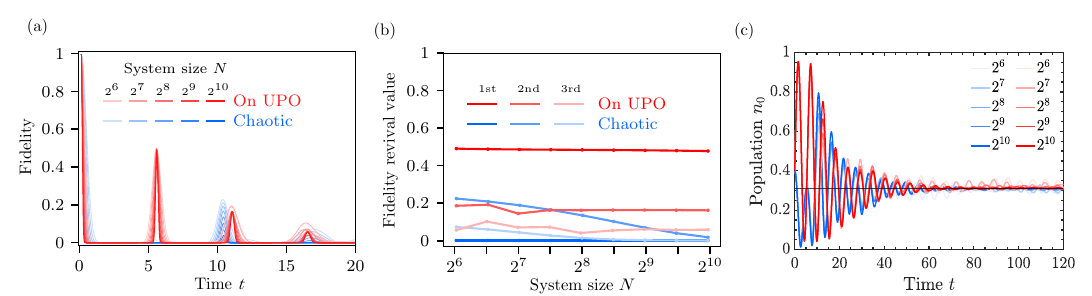}
    \caption{(a) The survival probability of an initial state on UPO (red) and off UPO, i.e.,~on a chaotic trajectory (blue). (b) The first three revival amplitudes for dynamics starting on UPO (red) and off UPO (blue). The dynamics started on UPO has revivals robust to increasing the system size, whereas the dynamics off UPO does not show revivals in the thermodynamic limit. (c) The spin$-0$ population dynamics for two different initial states at the same energy where one is chosen on the UPO (red) $\ket{\zeta_s}\equiv\ket{n_0,m,\theta,\eta}=\ket{0.4,0,\pi,0}$ and the other off UPO on a chaotic trajectory (blue) $\ket{\zeta_c}=\ket{0.4,0,0,\pi}$. The saturation values denote the system size given in the legend. The black solid line is the microcanonical ensemble prediction.}
	\label{fig:spectral_correlation}
\end{figure*}

Furthermore, we present the fidelity and local observable dynamics initiated with two coherent states in \autoref{fig:spectral_correlation} where (i) $\ket{\zeta_s}=\ket{n_0,m,\theta,\eta}=\ket{0.4,0,\pi,0}$ is on the UPO at $E_n/N=0.24$ demonstrating robust revivals in fidelity $F=\vert\langle \zeta(0)\vert \zeta(t)\rangle \vert^2$ at the UPO periods in (a,b) and (ii) $\ket{\zeta_c}=\ket{0.4,0,0,\pi}$ is on a chaotic trajectory and hence leading to a fidelity that does not revive. \autoref{fig:spectral_correlation}(c) shows how the dynamics of spin$-0$ population exhibit oscillations with the same frequency of the UPO at energy density $E=0.24$. These oscillations persist as we increase the atom number in the condensate demonstrating the robustness of scar dynamics. Eventually the oscillations decay to a value over time which is predicted by the microcanonical ensemble of states taken in a narrow energy window. Therefore, the observable thermalizes regardless of initiating the dynamics with an initial state on UPO (red) or off UPO (blue). This behavior is consistent with the entanglement entropy and eigenstate expectation values in \autoref{fig:chaos_diagnosis}(a)-(b). Compared to lower $p=0.05$ \cite{evrard-QuantumScars-2024}, we observe significantly larger amplitudes in fidelity revivals and local observable oscillations at $p=0.5$. We ergo conclude that increasing the integrability breaking strength $p$ greatly enhances the revivals and oscillations due to scarring, and allows for higher harmonics to manifest in the fidelity. Interestingly and counter-intuitively, moving away from the integrable point $p=0$ does not only render the entire spectrum more chaotic but also strengthens the signatures of the quantum scars.

\subsection{Shortest periodic orbit period}

As will be explained in the next section, antiscarring requires knowledge of the shortest periodic orbit period $t^*$. Although this is difficult to definitively answer, we already analytically know a lower bound to the periods of the UPOs that reside on the plane $m=\eta=0$, \autoref{fig:upo_period&poincare}(b). Here, to estimate the time $t^*$, we utilize two spectral functions, namely spectral rigidity $\Delta_3$ \cite{dyson1963statistical,guhr1998random} and SFF \cite{mehta2004random}. 

Spectral rigidity is expected to increase logarithmically in the energy window size $\Delta E$ and saturate around $1/t^*$ \cite{berry1985semiclassical,d2016quantum}. We compute the behavior of $\Delta_3$ for \autoref{eq.Hamiltonian2ndQuant} at three different condensate sizes $N$ as shown in \autoref{fig:upo_period&poincare}(c). As expected, we can identify a distinct logarithmic scaling in energy interval, and a plateau is reached around $\Delta E \sim 2\pi/T_{\text{UPO}}^*$ for $N=200$, where $T_{\text{UPO}}^*= \sqrt{2}\pi \simeq 4.44$ is the shortest UPO period among the UPOs defined on the $(n_0,\theta)$ plane. This observation implies that the shortest UPO might actually be the shortest periodic orbit in the entire phase space, i.e., $T_{\text{UPO}}^*=t^*$ (see \appref{appendixSec:unfolding} for the details of the rigidity calculation). 

Another important spectral correlation function is the \textit{spectral form factor} (SFF). The SFF is defined as the Fourier transform of the two-point spectral correlation and for discrete spectrum it reads,
%%%%%%%%%%%%%%%%%%%
\begin{equation}
K(t) = \left\langle \left|\sum_{n=1}^D e^{iE_n t}\right|^2\right\rangle, 
\end{equation}
%%%%%%%%%%%%%%%%%%%
where $D$ is the Hilbert space dimension, and $\langle \cdots \rangle$ refers to disorder averaging. In a generic quantum chaotic system, the SFF shows a `dip-ramp-plateau' behavior \cite{mehta2004random,haake2010quantum}. The `ramp' and `plateau' match the RMT prediction, while the `dip' part visible in early time is considered non-universal. It has been established~\cite{eichmann-ScaledEnergySpectroscopy-1988} that the connected SFF of quantum systems with a semiclassical limit,
\begin{eqnarray}
K_c(t) = \frac{1}{D} \left( \bigg \langle \bigg \vert\sum_{n=1}^D e^{iE_n t} \bigg \vert^2 \bigg \rangle - \bigg \vert \bigg \langle \sum_{n=1}^D e^{iE_n t} \bigg \rangle \bigg \vert^2 \right),
\end{eqnarray}
is strongly peaked at the periods of the POs, and weighted by the stability of the orbits. The peak at $t^*$ indicates the breakdown of RMT for longer range energy correlations than $\Delta E > 2\pi/t^*$. Since SFF is not self-averaging, we simulate a series of statistically similar systems with coupling $p\in [0.5-\frac{0.1}{N},0.5+\frac{0.1}{N}]$ such that the semiclassical limit at $p=0.5$ is restored as $N\rightarrow \infty$. \autoref{fig:upo_period&poincare}(d) shows that $K_c(t)$ follows GOE (dashed-black) in late times, whereas it is peaked around the shortest UPO period $T_{\text{UPO}}^*$ in early times (inset), agreeing well with $\Delta_3$. Here we normalize cSFF by $D$ for scaling purposes such that for different system sizes $N$ cSFFs overlap with each other approximately. Further details on the SFF calculation are provided in \appref{appendixSec:SFF}.

\section{Antiscarring}

Let us define a stack of eigenstates within an energy window $E \in \Delta E$ with the center of energy window being $E_0$ as $\rho = \sum_{E_n \in \Delta E} \vert \psi_n \rangle \langle \psi_n \vert f(E_n) $, where $f(E_n)$ is a filter function, e.g.,~ a Gaussian or box filter. We project the stack $\rho$ on a measure state $\ket{\zeta}$ of energy  $E_0 =\bra{\zeta} \hat H \ket{\zeta}$, translating 
%%%%%%%%%%%%
\begin{align} \label{eq:stacking_theorem}
 \mathcal{S}_{\ket{\zeta}} &= \text{Tr} \lbrace \vert \zeta \rangle \langle \zeta \vert \rho\rbrace =  \sum_{E_n \in \Delta E}\left|\braket{\zeta}{\psi_n} \right|^2 f(E_n) \nonumber \\
& = \int_{E \in \Delta E} dE \; f(E) \left\langle \zeta \middle| \delta(E-\hat H) \middle |\zeta \right\rangle \nonumber \\
& = \frac{1}{2\pi} \int_{E \in \Delta E} dE \; f(E)\left\langle \zeta \; \middle| \int_{-\infty}^{+\infty} dt \; e^{i(E-\hat{H})t} \middle |\; \zeta  \right\rangle \nonumber \\
& =   \frac{1}{2\pi}  \int_{-\infty}^{+\infty} dt \mathcal{A}_{\zeta}(t) \Omega(t)
\end{align}
%%%%%%%%%%%%
where we define a temporal cutoff function $\Omega(t) \equiv \int_{E \in \Delta} dE  \; f(E) e^{i(E-E_0)t}$ and $\mathcal{A}_{\zeta}(t)=\braket{\zeta}{\zeta(t)} e^{iE_0 t} = \langle \zeta \vert e^{-i(\hat H-E_0) t}\vert \zeta \rangle$ is the survival probability amplitude of the measure state $\vert \zeta \rangle$. If $f(E)$ is bounded with width $\Delta E$, $\Omega(t)$ which is the Fourier transform of $f(E)$ around $E_0$, would as well be bounded with the width $\Delta T \sim 2\pi/\Delta E$ according to energy-time uncertainty principle, therefore acting as a temporal cutoff function. 

For the chaotic spinor condensate defined in Eq.~\eqref{eq.Hamiltonian2ndQuant}, we take the measure states to be coherent states that span a phase space in the semiclassical limit $N\rightarrow \infty$. First, we note the importance of the energy criterion for the measure state $\ket{\zeta}$: The semiclassical limit of Eq.~\eqref{eq.Hamiltonian2ndQuant} has a continuous family of UPOs in the energy density between $-0.25 < E/N < 0.75$, and at the center of spectrum where the density of states peak, the periods of these UPOs are sufficiently short to scar the eigenstates. Therefore, the energy criterion of the measure states is essential to suppress the effect of scarring by UPOs at energies different than $E_0$. In fact, stacking eigenstates scarred by multiple UPOs for any coherent state would result in a strongly scarred stack instead of uniformity (not shown here). 

\begin{figure*}[htb]
\centering
\includegraphics[width=\textwidth]{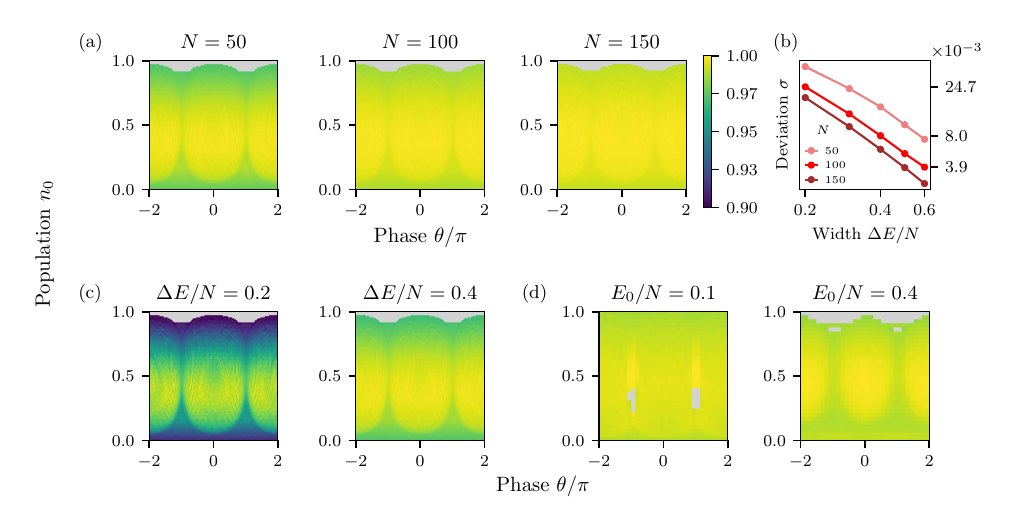}
\caption{(a) Projected eigenstate stacking at energy density center $E_0/N=0.24$ with energy window fixed $\Delta E/N=0.6$ with increased uniformity observed with increasing $N$. (b) The deviation from perfect uniformity $\sigma$ with respect to energy window for different atom numbers and for different system sizes. (c) Projected eigenstate stacking at energy density center $E_0/N=0.24$ with atom number fixed to $N=100$ and for smaller energy windows than $\Delta E/N=0.6$, showing more inhomogeneity. (d) Projected eigenstate stacking with atom number fixed to $N=100$ and energy window $\Delta E/N=0.6$ with different energy density center $E_0/N$. Approximate uniformity at finite-sizes persists. All phase space diagrams in (c) and (d) share the same colorbar as in (a).}
\label{fig:stack_scaling}
\end{figure*}
Then it is straightforward to see that $\mathcal{S}_{\ket{\zeta}} = \mathcal{S}, \forall \ket{\zeta}$ must hold at energy ${E_0}$, as long as $\mathcal{A}_{\zeta}(t<t^*) = \mathcal{A}(t<t^*)$ for the shortest periodic orbit period $t^*$ \cite{keski-rahkonen-AntiscarringChaotic-2024}. For coherent states, even at the same energy, we have $\mathcal{A}_{\zeta}(t<t^*) = \exp(-b(\zeta) t^2)$. Hence the decay exponent $b(\zeta)$ depends on the choice of coherent states (see \appref{AppendixSec:coherent}), rendering our measure states imperfect from the original perspective of the theorem, assuming identical dispersion for the measure states \cite{keski-rahkonen-AntiscarringChaotic-2024}. Consequently, such imperfect measure states would instead lead to \textit{approximate uniformity}, achieving perfect uniformity only in the thermodynamic limit as $\Delta T \rightarrow 0$ due to extensive energy window $\Delta E \propto N$.

Given that $\mathcal{S}_{\ket{\zeta}}=\sum_{E_n \in \Delta E} Q_n (\zeta)  f(E_n)$, we visualize the stacking by computing the equal-energy projection function
%%%%%%%%%%%%%%%%%%%%%%
\begin{eqnarray}
\mathcal{P}^{E_0}_n\left(n_0, \theta\right)&=& \mathcal{P}_n \left(n_0, \theta\right) \hspace{1mm}\delta\left(E_0-\langle\zeta| \hat{H}|\zeta\rangle\right),\label{eq:projectionF}
\end{eqnarray}
%%%%%%%%%%%%%%%%%%%%%%
where $\mathcal{P}_n\left(n_0, \theta\right)=\frac{1}{d\left(n_0, \theta\right)} \iint
d m d \eta \hspace{1mm} Q_n \left(n_0, \theta, m, \eta\right)$ is the projection function to detect scarring \cite{evrard-QuantumScars-2024}, and $d\left(n_0, \theta\right)=\iint d m d \eta \hspace{1mm} \delta\left(E_0-\langle\zeta| \hat{H}|\zeta\rangle\right)$ is the density of states at energy $E_0$. Eq.~\eqref{eq:projectionF} leads to the projected stacking,
\begin{equation}\label{eq:proj_stack}
\mathcal{S}\left(n_0, \theta\right) = \sum_{E_n \in \Delta E} P_n^{E_0} \left(n_0, \theta\right) f(E_n),
\end{equation}
where $f(E)$ is taken to be a Gaussian distribution $f(E) \propto \exp(-(E-E_0)^2/2(\Delta E/2)^2)$ with $E_0/N=0.24$ and twice of the standard deviation, $\Delta E$, of $f(E)$ effectively serving as the energy window width. 

\autoref{fig:stack_scaling}(a) and (c) show projected eigenstate stacking for different system sizes in fixed energy window $\Delta E/N=0.6$ and for different energy windows at fixed system size $N=100$, respectively. 
We observe that the stacking becomes more uniform as the energy windows becomes wider and the system size increases. We quantify the uniformity with the standard deviation of the projected stacking distribution, $\mathcal{S}\left(n_0, \theta\right)$ for all $(n_0,\theta)$, which decays as a power-law in the energy window size (\autoref{fig:stack_scaling}(b)). Furthermore, the deviation decreases with increasing atom number $N$. Therefore, we see that the phase space approaches the perfect uniformity in the thermodynamic limit. In the thermodynamic limit, i.e.,~the semiclassical limit, $N \rightarrow \infty$, $\Delta E \rightarrow \infty$, and therefore the cutoff time $\Delta T \rightarrow 0$, the differences in the decay of coherent states vanish. However, the stacking window still remains as a small portion of the full spectrum. The eigenstate stacking theorem also holds at different energy density centers. \autoref{fig:stack_scaling}(d) shows results for $E_0/N=0.1,0.4$ at system size $N=100$ and energy window $\Delta E / N =0.6$, demonstrating near-uniform behavior across all cases studied.

Although the stacking is approximately uniform for finite-size condensates, we still observe the antiscarring of an ensemble of eigenstates, i.e.,~a suppression of the probability density around the scarring UPO. In \autoref{fig:scar_antiscar}(b), we introduce a cumulative scar, defined as the sum of projected eigenstates in the energy window, formally $\sum_{\mathcal{F}_n > 0.03,E_n \in \Delta E} \mathcal{P}_n^{E_0}(n_0,\theta) f(E_n)$. In this construction, we select only the states that are strongly scarred by the UPO at $E_0/N=0.24$ according to the scarmometer $\mathcal{F}_n > 0.03$ in \autoref{fig:scar_antiscar}(a).  On the other hand, as predicted by the stacking theorem, the projection of the rest of the eigenstates $\sum_{\mathcal{F}_n < 0.03, E_n \in \Delta E} \mathcal{P}_n^{E_0}(n_0,\theta)f(E_n)$ exhibits a diminished probability density around the same UPO in \autoref{fig:scar_antiscar}(c), which is highlighted with white-dashed line. Therefore, we can conclude that the stacking windows to observe uniformity in \autoref{fig:stack_scaling}(c) are larger than the energy scale of the shortest periodic orbit, $\Delta E > 2\pi/t^*$, as required by the original theorem~\cite{keski-rahkonen-AntiscarringChaotic-2024}.

%%%%%%%%%%%%%%%%%%%%%%%%%%%%%%%%%%%%%%%%%%%%%%%%%%%%
%%%%%%%%%%%%%%%%%%%%%%%%%%%%%%%%%%%%%%%%%%%%%%%%%%%%

\section{Discussion and outlook}

We unveiled an unknown facet of scarring in quantum systems with many particles -- antiscarring, referring to the suppressed probability density of a set of eigenstates along the scar-generating UPO to ensure the uniformity of the whole in the underlying phase space. The uniformity was enforced by the eigenstate stacking theorem, which we extended to systems with any measure basis by loosening an assumption required in the more strict version of the theorem~\cite{keski-rahkonen-AntiscarringChaotic-2024}. 

The chaotic spinor condensates are many-body quantum gases with collective interactions \cite{evrard-QuantumScars-2024,austin2024observation}, hence they possess a semiclassical limit, which presents a natural and optimal choice for measure basis. For many-body systems that do not have a semiclassical limit, one can still define effective phase spaces through various methods \cite{ho2019pxp,Hallam2023,evrard-QuantumManybody-2024,pizzi2024scar} and check for the applicability of the stacking theorem together with the presence of antiscarring. Given that the SU(3) coherent states can be prepared in the laboratory \cite{PhysRevA.103.L031302, austin2024observation}, the eigenstate stacks can be stated in time as in Eq.~\eqref{eq:stacking_theorem}, and the uniformity of the phase space is expected only in early times, our theory is in experimental reach. Specifically, via the phase space projections based on the coherent states, one can directly probe the (approximate) uniformity of eigenstate stacking theorem and scars, which then leads to a direct probe of antiscarring. 
On a different note, quantum scarring can be detected by probing the revivals in the time evolution of a coherent state prepared on a UPO. In fact, we showed that a spinor condensate with a larger integrability breaking is more chaotic while exhibiting a stronger dynamical signature of scarring simultaneously. Antiscarring is not limited to quantum systems with UPOs: as the stacking theorem also holds for stable POs, we predict that many-body systems exhibiting weak ergodicity breaking \cite{Serbyn_2021} must also show a form of antiscarring.

Even though our findings suggest ways to exploit these quantum interference deviations from naive ergodicity in experiments and devices, we do not suspect any gross violations of thermodynamics. Nevertheless, we expect that our work will inspire future investigations into scarring and the quantum nature of ergodicity, particularly paving the way for the experimental observation of this previously hidden aspect of quantum scarring. For example, whether antiscarring has an independent dynamical signature in observables or fidelity, such as a slow decay rate in open systems~\cite{kaplan-ScarAntiscar-1999}, and if so, which initial states can lead to antiscarring dynamics, are interesting questions to answer in the future.

\section{Acknowledgements}

The authors thank L. Kaplan, B. Evrard and A. Pizzi for stimulating discussions. 
A.M.G. thanks the Studienstiftung des Deutschen Volkes PhD Fellowship and the  Harvard Quantum Initiative for financial support. J.K.-R. thanks the Oskar Huttunen Foundation for the financial support. C.B.D was supported by the ITAMP grant No.~2116679 and F100 initiative at Indiana University at Bloomington. This project was also supported by the National Science Foundation (Grant No.~2403491). 

\appendix

\section{\label{AppendixSec:UPOPeriod}Derivation of Eq.~\eqref{eq:TUPO}}

From Eqs.~\eqref{eq:UPOenergy} and~\eqref{eq:EOMUPO},  we derive the period of the UPOs, $T_{\rm UPO}$ as,
\begin{equation}\label{eq:upo_periodicity}
    T_{\rm UPO} = \int_{n_{-}}^{n_{+}} \frac{2 dn}{\sqrt{(4p^2+2E)n(1-n)-E^2-\left( n(1-n) \right)^2}}
\end{equation}
where $n_{\pm} = \frac{1}{2} \left(1\pm \sqrt{1-4\left( \sqrt{p^2+E} - p \right)^2} \right)$. Let $x=n-1/2$, then the integral above becomes,
\begin{equation}
    I = \int_{-x_{m}}^{x_{m}} \frac{dx}{\sqrt{-x^4+bx^2+c}},
\end{equation}
where $b=-2(2p^2+E-1/4)$, $c= p^2-(E-1/4)^2$ and $x_{m}= \frac{1}{2}\sqrt{1-4\left( \sqrt{p^2+E} - p \right)^2}$. We can factorize the denominator of the integrand, 
\begin{equation}
\frac{1}{\sqrt{-x^4+bx^2+c}} = \frac{1}{\sqrt{-(y_1 -x^2)(y_2 -x^2)}}
\end{equation}
with $y_{1,2} = -(2p^2+E-\frac{1}{4})\pm 2p \sqrt{p^2+E}$ and we note $y_1= x_{m}^2$. The change of variables \( x = x_m \cos \theta \) transforms the integral into a form involving a complete elliptic integral of the first kind,
\begin{equation}
\begin{aligned}
I &= \frac{2}{\sqrt{y_1-y_2}} K\left(\sqrt{\frac{y_1}{y_1-y_2}}\right).
\end{aligned}
\end{equation}
For $p=0.5$, $y_1>0$ and $y_2 < 0$ hold. This completes the derivation of Eq.~\eqref{eq:TUPO} following $T_{\text{UPO}} = 2 I$.

\section{Detailed proof of Eq.~\eqref{eq:stacking_theorem}}
Here we demonstrate the detailed proof of the stacking theorem. Utilizing the property of Delta-Dirac function,
\begin{widetext}
\begin{eqnarray}
\int_{E \in \Delta E} dE \;\left|\braket{\zeta}{E_k}\right|^2 \delta(E-E_k) f(E) \; = 
\left\{
\begin{aligned}
 &\left|\braket{\zeta}{E_k} \right|^2 f(E_k)   &E_k \in \Delta E \\
&0   &E_k \notin \Delta E 
\end{aligned}
\right.
\end{eqnarray}
\end{widetext}
we rewrite the stacking function as
\begin{eqnarray}
\mathcal{S}_{\ket{\zeta}} &=& \sum_{E_n \in \Delta E}\left|\braket{\zeta}{E_n} \right|^2 f(E_n) \notag \\
&=& \sum_{k=1}^{D} \int_{E \in \Delta E} dE \; \left|\braket{\zeta}{E_k}\right|^2 \delta(E-E_k) f(E), \notag\\
&=& \int_{E \in \Delta E} dE \;f(E)\left(\sum_{k=1}^{D} \left|\braket{\zeta}{E_k}\right|^2 \delta(E-E_k) \right) \notag\\
& =& \int_{E \in \Delta E} dE \; f(E) \left\langle \zeta \middle| \sum_{k=1}^{D}\ket{E_k}\bra{E_k}\delta(E-E_k) \middle |\zeta \right\rangle \notag\\
& =& \int_{E \in \Delta E} dE \; f(E) \left\langle \zeta \middle| \delta(E-\hat H) \middle |\zeta \right\rangle \notag\\
& =& \frac{1}{2\pi}\int_{E \in \Delta E} dE \;  f(E)\left\langle \zeta \; \middle| \int_{-\infty}^{+\infty} dt \; e^{i(E-\hat H)t} \middle |\; \zeta  \right\rangle, \notag
\end{eqnarray}
where in the last equality we use another definition for the Delta-Dirac function $\delta(E-\hat H) = \frac{1}{2\pi} \int_{-\infty}^{+\infty} dt \; e^{i(E-\hat H)t}$. This proves Eq.~\eqref{eq:stacking_theorem}.

\section{\label{AppendixSec:coherent}Decay properties of coherent states}

\begin{figure}[htp]
	\includegraphics[width=1\columnwidth]{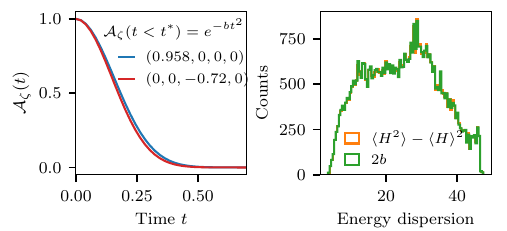}
	\caption{\textit{Left:} Survival probability amplitude $\mathcal{A}_{\zeta}(t) = \braket{\zeta}{\zeta(t)} e^{iE_0 t}$ at $t\lesssim t*$ of two coherent state with $E_0/N \simeq 0.24$. The coherent state marked by the blue curve is released on the highest point (maximal $n_0$) on the $E=0.24$ UPO at $t=0$. It is postulated that the shape of $\mathcal{A}_{\zeta}(t)$ follows $\exp(-bt^2)$.  \textit{Right:} Histograms of $2b$ and $\langle \Delta H^2 \rangle$ of all coherent states around energy shell $E=0.24$ with width $\epsilon = 0.03$ which represents the chosen criterion for equal energy in our numerical simulations. Parameter $b$ is obtained by fitting $\mathcal{A}_{\zeta}(t)$ at short time with the hypothesized Gaussian. The two histograms match perfectly which indicates the correctness of the hypothesized Gaussian shape.}
	\label{fig:gaussian_decay}
\end{figure}

The eigenstate stacking is closely related to the early-time decay properties of the measure states. Choosing the coherent states as the natural basis of measure, here we analyze their decay properties.
For coherent states $\ket{\zeta}$ we observe the survival probability amplitude $\mathcal{A}_{\zeta}(t\lesssim t*) = \exp(-bt^2)$ is real when $t\lesssim t*$ . We fit the parameter $b$ for each coherent state on energy shell $E_0/N = \langle \hat{H} \rangle /N = 0.24$ and compare it with the energy dispersion $\langle \Delta \hat{H}^2 \rangle= \langle \hat{H}^2\rangle - \langle \hat{H} \rangle ^2$, finding that $\langle \Delta \hat{H}^2 \rangle$ agrees perfectly with $2b$, see \autoref{fig:gaussian_decay}. Indeed, this can be shown for $\mathcal{A}_{\zeta}(t)$ when $t \rightarrow 0$ that $\mathcal{A}_{\zeta}(t) \simeq (1 - i \langle \hat{H} \rangle t -\frac{1}{2} \langle \hat{H}^2 \rangle t^2) (1+iE_0t -\frac{1}{2}E_0^2t^2) \simeq 1-\frac{1}{2}(\langle \hat{H}^2 - E_0^2 \rangle)t^2 \simeq \exp(-\langle \Delta \hat{H}^2 \rangle t^2/2)$.
%
%%%
%%%

The right of \autoref{fig:gaussian_decay} also shows that the decay rates of coherent states under the Hamiltonian evolution differ significantly from each other, showing why we observe only approximate uniformity in phase space for finite-size condensates.

%%%%%%%%%%%%%%%%%%%%%%%%%%%%%%%%%%%%%%%%%%%%%%%%%%%%%%%%%%%%%%%%%%%%%

\section{\label{appendixSec:unfolding} Unfolding procedure and spectral properties}

\begin{figure}[t!]
    \includegraphics[width=0.47\columnwidth]{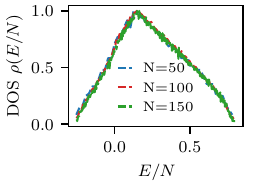}
    \includegraphics[width=0.47\columnwidth]{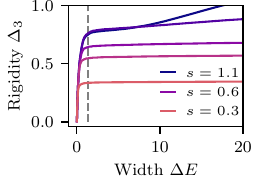}
    \caption{\textit{Left:} Normalized spectral density of the Hamiltonian at $p=0.5$ for size $N=50,100,150$ computed with exact diagonalization. A peak appears in the spectrum around $E/N=0.15$. \textit{Right:} Rigidity $\Delta_3$ for different smoothing factor $s$ for system size $N=150$. Large $s$ leads to underfitting (e.g. $s=1.1$) while small $s$ leads to overfitting. The dashed gray line marks $\Delta E_c = 2\pi /T_{\text{UPO}^*}$, i.e. the inverse of shortest period of the UPOs on $m=\eta=0$ plane, which serves as a upper bound of shortest PO in the entire system, meaning that $\Delta_3$ should develop a plateau only when $\Delta E > \Delta E_c$. For this reason, $s\le 0.3$ is likely to be overfitted, and we pick $s=0.6$ as a proper choice.}
    \label{fig:dos&unfold}
\end{figure}

In quantum systems, analyzing the statistical properties of energy levels requires separating global trends from local fluctuations, e.g.,~the spectrum of the system at $p=0.5$ in \autoref{fig:dos&unfold} has a global shape. This process, known as \textit{unfolding}, normalizes the energy spectrum to facilitate meaningful statistical analyses by ensuring a uniform average spacing between energy levels. Through the unfolding process, the spectral distribution is adjusted to achieve a uniform density, effectively removing the influence of the global trend of the original spectrum, which enables meaningful comparison. 
The unfolding begins with constructing the cumulative spectral density (staircase function) $N_{\text{emp}}(E)$, which counts the number of energy levels $E_i$ up to a certain energy $E$:
%%%%%%%%%%%%%%%%%%%%%%%
\begin{equation}
N(E) = \sum_{i=1}^{N} \Theta(E - E_i)
\end{equation}
%%%%%%%%%%%%%%%%%%%%%%%
where $\Theta(x)$ is the Heaviside step function. 
%
%%
%%%
%%%
%%
%
For identifying the global trend in $N(E)$, we apply a smooth cubic spline function, labeled as $\xi$, to fit the density of states. This fitting yields $N_{\text{sm}}(E) = \xi(E, N(E), s)$, where the smoothness parameter $s$ manages the trade-off between data accuracy and spline smoothness. Each original energy level $E_i$ is then mapped to an unfolded level $\tilde{E}_i = N_{\text{sm}}(E_i)$, and the staircase function is viewed as a function of $\tilde{E}_i$, $N = N(\tilde{E}_i)$.

Spectral rigidity $\Delta_3$ \cite{dyson1963statistical,guhr1998random} quantifies the fluctuation (or rigidity) of the spectrum at different length scales by measuring the deviation of the cumulative spectral density from a linear fit over an interval of length $\ell$, defined as
%%%%%%%%%%%%%%%%%%%%%%%%%%%%%
\begin{equation}
\Delta_3(\ell) = \frac{1}{\ell} \min_{A, B} \int_{\tilde{E}_0}^{\tilde{E}_0+\ell} \left[ N(\tilde{E}) - A\tilde{E} - B \right]^2 d\tilde{E},
\end{equation}
-%%%%%%%%%%%%%%%%%%%%%%%%%%%%%
where $A$ and $B$ are fitting parameters, $\tilde{E}$ is the unfolded energy and $N(\tilde{E})$ is the unfolded cumulative spectral density. 

Berry~\cite{berry1985semiclassical} demonstrated that for classically chaotic systems, the \textit{spectral rigidity} \( \Delta_3 \) exhibits logarithmic dependence on the energy interval length \( \ell \) as predicted by RMT, while for integrable systems it grows linearly. His argument also revealed a limitation of RMT: the correspondence between chaotic systems and RMT holds only up to a maximum \( \ell \) value \( \ell_{\text{max}} \), determined by the shortest classical periodic orbit \( T_{\text{min}} = h/\ell_{\text{max}} \). Beyond this threshold, \( \Delta_3 \) saturates for realistic quantum systems. The same saturation was also observed in various systems, for instance integrable \cite{casati1985energy}, chaotic (e.g.,~billiard \cite{sieber-NongenericSpectral-1993}), transition system \cite{seligman-QuantumSpectra-1984,seligman-FluctuationsQuantum-1985} and even random-matrix ensembles \cite{prakash-SaturationNumber-2016}. 

The computation of spectral rigidity in practice, however, relies on how the unfolding is conducted, while the latter generally does not have a perfect answer. A necessary condition for an ideal fit of $N_{\text{sm}}(E)$ ensures that the unfolded spectrum $\{ \tilde{E}_i \}$ has an average level spacing of unity. In practice, we adjust $s$ such that it ensures the unfolded average spacing $\tilde{\delta}$ is very close to unity, as well as ensuring the plateau of spectral rigidity $\Delta_3$ at large window $\ell$ can be observed, avoiding both underfitting and overfitting. The effect of varying $s$ is shown in \autoref{fig:dos&unfold}, from which we choose $s$ around $0.6$ as is shown in \autoref{fig:upo_period&poincare}(c). Note that in demonstrations we rescale $\ell$ to $\Delta E =\ell \langle d \rangle$ where $\langle d \rangle$ is the mean level spacing of the original spectrum, for easier comparison with PO periodicity.

%%%%%%%%%%%%%%%%%%%%%%%%%%%%%%%%%%%%%%%%%%%%%%%%%%%%%%%%%%%%%%%%%%%%%%%%%%%%%%%%%%%%%%%%%%%%
\begin{figure}[htp]
	\includegraphics[width=1\columnwidth]{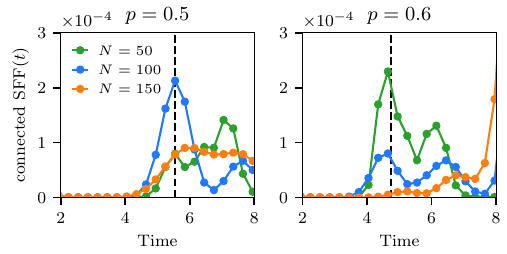}
	\caption{cSFF with a filter function that suppresses contributions of levels far away from $\epsilon_0=0.24$ for $p=0.5$ and $0.6$. A peak is observed at finite size $N$ which matches $T_{\text{UPO} \epsilon_0=0.24} \simeq 5.6$ (dashed black line). However, the height of the peak does not grow monotonously or remain the same with $N$ in contrast to the SFF with no filter function.}
\label{fig:sff_app}
\end{figure}
\section{\label{appendixSec:SFF}Spectral form factor}

Here we briefly review the Gutzwiller trace formula. The semiclassical spectral theory expresses the fluctuations in the density of states in terms of sums over classical periodic orbits, i.e.,~the Gutzwiller trace formula \cite{gutzwiller1971periodic,Heller_book_2018},
%%%%%%%%%%%%%%%%%%%
\begin{equation}
\rho_{\text{osc}}(E)=\sum_n A_n \exp \left[\frac{i}{\hbar} S_n(E)\right]
\end{equation}
%%%%%%%%%%%%%%%%%%%
where $S_n(E)$ is the classical action of the n$^{\text{th}}$ PO and amplitudes $A_n$ is proportional to the inverse square root of stability exponents.
%%%%%%%%%%%%%%%%%%%
Using the semiclassical expression, the semiclassical SFF can be derived as \cite{Stockmann_1999},
\begin{eqnarray} \label{eq:sff_semi}
K_c(t)&=&\sum_{n, m}{ }^{\prime} A_n A_m^*\left\langle\exp \left[\frac{i}{\hbar}\left(S_n-S_m\right)\right]\right\rangle \\
&\times &\delta\left(t-\frac{t_n+t_m}{2}\right) \notag
\end{eqnarray}
%%%%%%%%%%%%%%%%%%%
If one only considers the diagonal term, the above equation becomes,
%%%%%%%%%%%%%%%%%%%
\begin{equation}
K_D(t)=g \sum_n\left|A_n\right|^2 \delta\left(t-t_n\right),
\end{equation}
%%%%%%%%%%%%%%%%%%%
This semiclassical approximation of SFF suggests a peak at $t=t_n$ due to the contribution of the n$^{\text{th}}$ PO.

In the main text, we demonstrated cSFF is peaked around the period of the shortest UPO. We further investigate the imprints of POs on the cSFF by using a filter function $f(E) = e^{-a(E/N-\epsilon_0)^2}$ where $\epsilon_0$ is the energy density of a certain UPO we focus on. The filtered cSFF becomes,
%%%%%%%%%%%%%%%%%%%
\begin{equation}
\Bar{K}_c (t) = \left\langle \left|\sum_{n=1}^D f(E_n)e^{iE_n t}\right|^2\right\rangle - \left|\left\langle\sum_{n=1}^D f(E_n)e^{iE_n t}\right\rangle \right|^2
\end{equation}
%%%%%%%%%%%%%%%%%%%
With the filter function, the contribution of levels far away from $\epsilon_0$ is exponentially suppressed. In \autoref{fig:sff_app} we show the filtered cSFF with system sizes $N=50,100,150$ for two different coupling strength $p=0.5,0.6$. We alter $a$ with $N$ such that $f(E)$ approximately focus on the same amount of levels around $\epsilon_0=0.24$. We notice the peak at the period of the UPO with energy $\epsilon_0=0.24$, however the height of the peak does not increase monotonically with $N$. This could be due to the fact that the peak caused by the PO in Eq.~\eqref{eq:sff_semi} is inversely proportional to its stability exponent, and hence the contribution from an unstable PO in the finite-size SFF will be suppressed as system size increases towards the semiclassical limit.

%\bibliography{References}

%Control: production of eprint (1) enabled
%

\end{document}